\documentclass[]{aastex631}

\shorttitle{The Galileo Project}
\graphicspath{{./}{figures/}}

\begin{document}

\title{Overview of The Galileo Project}

\author{Abraham (Avi) Loeb} 
\affiliation{Head of the Galileo Project, Astronomy Department, Harvard University\\60 Garden Street, Cambridge, MA 02138, USA\\}
\author{Frank H. Laukien}
\affiliation{Co-Founder of the Galileo Project, Department of Chemistry and Chemical Biology, Harvard University\\
12 Oxford Street, Cambridge, MA 02138, USA\\}







\begin{abstract}
The Galileo Project is the first systematic scientific research program in search for potential astro-archaeological artifacts or remnants of extraterrestrial technological civilizations (ETCs) or potentially active equipment near Earth. Taking a path not taken, it conceivably may pick some low-hanging fruit, and without asserting probabilities - make discoveries of ETC-related objects, which would have far-reaching implications for science and our worldview. \end{abstract}


\section{Introduction}

The Galileo Project~\citep{thegalileoproject} is a scientific search program for potential astro-archaeological artefacts or remnants of extraterrestrial technological civilizations (ETCs), or potentially active extraterrestrial equipment near Earth. We co-founded the Project in July 2021. The Project's name was inspired by Galileo Galilei's
legacy of finding answers to fundamental questions by looking through new telescopes. The search is agnostic to the outcome. It represents a scientifically rigorous search for ETC artifacts, remnants, space trash or active equipment in the form of conceivable ETC-generated interstellar objects (ISOs), even if these ETCs may be extinct by now, or of potential ETC-generated or actively controlled Unidentified Aerial Phenomena (UAPs). The search could result in a mixed bag containing primarily (after the elimination of instrumental artifacts):

\begin{enumerate}
    \item {\bf Natural objects or Phenomena}, like: bugs, flying insects, birds or mammals, dust on lenses or windows, comets, asteroids, rocky meteors or atmospheric phenomena, including reflections or mirages, lightnings or plasma effects.
    \item {\bf Human-made objects or phenomena}, like: weather balloons, drones, helicopters, airplanes, rockets, spacecraft or satellites or human-made mirages.  
\end{enumerate}

Assembling high-quality traceable, scientific quality, simultaneously acquired multi-modal and multi-spectral data to rebut or mutually confirm anomalous experimental observations data on the first category, would be of interest to zoologists, atmospheric scientists and planetary scientists. The second category could be of interest to national security agencies. But anything else would be of great scientific interest to the Galileo Project. This third category includes objects that appear to be of artificial origin, e.g. showing extraterrestrial sensors, propulsion methods, heat or engine exhaust, landing or aerial maneuvering capabilities, or in the extreme - some nuts or bolts in high-resolution images of their surface, but moving or interacting in ways that cannot be reproduced by current human-made devices.  

The Galileo Project is a new scientific research initiative within the context of known physics. Its novel instruments will monitor the sky in the optical, infrared and radio bands, as well as in audio, magnetic field and energetic particles signals. The data will be analyzed by artificial intelligence algorithms that will aim to catalog objects or phenomena within the above-mentioned categories.

As Arthur Conan Doyle's fictional detective Sherlock Holmes noted: ``When you have eliminated all which is impossible, then whatever remains, however improbable, must be the truth.''~\citep{doyle2020adventure}.

\section{In Search for Technological Interstellar Objects}

Extraterrestrial equipment could arrive in two forms: defunct `space trash', similar to the way our own spacecraft will appear in a billion years, or functional equipment, such as an autonomous craft equipped by Artificial Intelligence (AI). The latter would be a natural choice for crossing the tens of thousands of light years that span the scale of the Milky Way galaxy and could exist even if the senders are not alive to transmit any detectable signals at this time. Hence, space archaeology for extraterrestrial equipment is a new observational frontier, not represented in the past history of the Search for Extraterrestrial Intelligence (SETI) which focused on contemporaneous electromagnetic signals and not long-lived physical objects which are gravitationally-bound to the Milky Way~\citep{lingam_loeb_2021}.

As an astronomer, one of us (A.L.) became interested in this subject after the observational discovery of interstellar objects~\citep{Loeb_book}. The first three interstellar objects were discovered only over the past decade (2014-2019). At the time of this writing, they include~\citep{SL22}:

\begin{enumerate}
    \item The first interstellar meteor, {\bf CNEOS 2014-01-08}, detected on January 8, 2014 by US Government sensors near Papua New Guinea~\citep{2019arXiv190407224S}. It was half a meter in size and exhibited material strength tougher than iron~\citep{2022RNAAS681S}. It was an outlier both in terms of its speed outside the Solar system (representing the fastest five percent in the velocity distribution of all stars in the vicinity of the sun) and its material strength (representing less than five percent of all space rocks). The Galileo Project plans an expedition to retrieve the fragments of this meteorite from the ocean floor in an attempt to determine the composition and potentially the structure of this unusual object and study whether it was natural or artificial in origin.
    \item The unusual interstellar object, {\bf `Oumuamua} (1I/2017 U1)~\citep{2021arXiv211015213L}, discovered by the Pan STARRS telescope in Hawaii on October 19, 2017, which was pushed away from the Sun by an excess force that declined inversely with distance squared~\citep{Micheli} but showed no evidence for cometary gases indicative of the rocket effect~\citep{Trilling}. 
    'Oumuamua had many other unusual features, such as an extreme (most likely flat) shape, high reflectivity, no jitter in its period of rotation as expected for cometary jets, and an origin in the Local Standard of Rest, consistent with a hypothetical planned mission to explore our inner habitable planets.     Another object, 2020 SO, exhibiting an excess push with no cometary tail, was discovered by the same telescope in September 2020. It was later identified as a rocket booster launched by NASA in 1966, being pushed by reflecting sunlight from its thin walls. The Galileo Project aims to design a space mission that will intercept or rendezvous with the next `Oumuamua and get high quality data that would allow to decipher its nature. The Project will also develop software that will identify targets of interest out of the data pipeline from the future Legacy Survey of Space and Time (LSST) of the Vera C. Rubin Observatory, when available.
    \item The interstellar comet, {\bf 2I/Borisov}~\citep{2021DPS5350502O}, was discovered on August 29, 2019 by the amateur astronomer, Gannadiy V. Borisov. This object resembled other comets found within the Solar system and was definitely natural in origin.
\end{enumerate}

It is intriguing that two out of the first three interstellar objects appear to be outliers relative to familiar Solar system asteroids or comets.

\section{Cosmic Perspective}

The chance of finding a civilization at exactly our technological phase is small, roughly one part in a hundred million — the ratio between the age of modern science and the age of the oldest stars in the Milky Way. Most likely, we would encounter extraterrestrial life or civilizations that are either way behind or way ahead of our scientific knowledge. To find the former class, we will need to visit oceans or the jungles of exo-planets, natural environments similar to those occupied by primitive human cultures over most of the past million years. This task would require a huge amount of effort and time given our current propulsion technologies. Chemical rockets take at least forty thousand years to reach the nearest star system, Alpha Centauri, which is about four light years away. Their speed is ten thousand times slower than the speed of light, implying a travel time of half a billion years across the Milky Way disk with over fifty  thousand light years in diameter.

But if the most advanced scientific civilizations started their scientific endeavor billions of years ago, we might not need to go anywhere since their equipment may have already arrived at our cosmic neighborhood in the form of interstellar artifacts. In that case, all we need to do is become curious observers of our skies and our immediate cosmic neighborhood in the solar system.

\section{A New Search}

The Galileo Project represents a new research initiative in astronomy. Existing astronomical observatories overwhelmingly target objects at great distances and have a limited field of view of the sky, whereas the Galileo Project aims to monitor the entire sky continuously and study fast-moving interstellar objects in the vicinity of Earth. It is an astronomy project since it analyzes data obtained by telescopes and searches for interstellar objects that could have originated outside the Solar system. In the second experimental track, the Project's novel observing strategy employs state-of-the-art cameras and computers that monitor the entire nearby sky in the optical, infrared and radio bands, as well as in audio, magnetic field and energetic particle signals.

Government agencies aim to protect the safety of military personnel and national security interests. From their perspective, reports by military staff members on Unidentified Aerial Phenomena (UAP), such as those documented by the National Intelligence Agencies and discussed during dedicated hearing sessions in the US Congress~\citep{odnireport}, are of primary importance for the first task, and data from military patrol sites are linked to the second objective. Government agencies have the responsibility to find out what the vast majority of UAP are, and for that purpose they must also attend to data of compromised quality such as blurry videos.

The task for scientists is complementary. Science does not need to explain most of the reports if their data is inconclusive. But even if only one object is of extraterrestrial technological origin among the clutter of many natural or human-made objects, it would represent the most consequential discovery in human history. To figure this out, scientists must have access to the highest quality data, such as a high-resolution image of an object showing unconventional propulsion, or a maneuver at extreme speeds, or exhibit kinematics outside the performance envelope of conventional aircraft and projectiles.

Moreover, scientists are concerned with all possible geographical locations even if they do not host military assets or national facilities. Extraterrestrial equipment might not adhere to national borders in much the same way that a biker navigating down the sidewalk does not care which of the possible pavement cracks is occupied by a colony of ants.

Satellite data may allow us to study UAP from above. This offers complementary opportunities to track their motion and image better than possible from ground-based ground based observatories. The Galileo Project is engaged in studying satellite data sets that are publicly available.

\section{Branches of Activity and Guiding Principles}

\begin{figure}[htb]
  \centering
  \includegraphics[width=\linewidth]{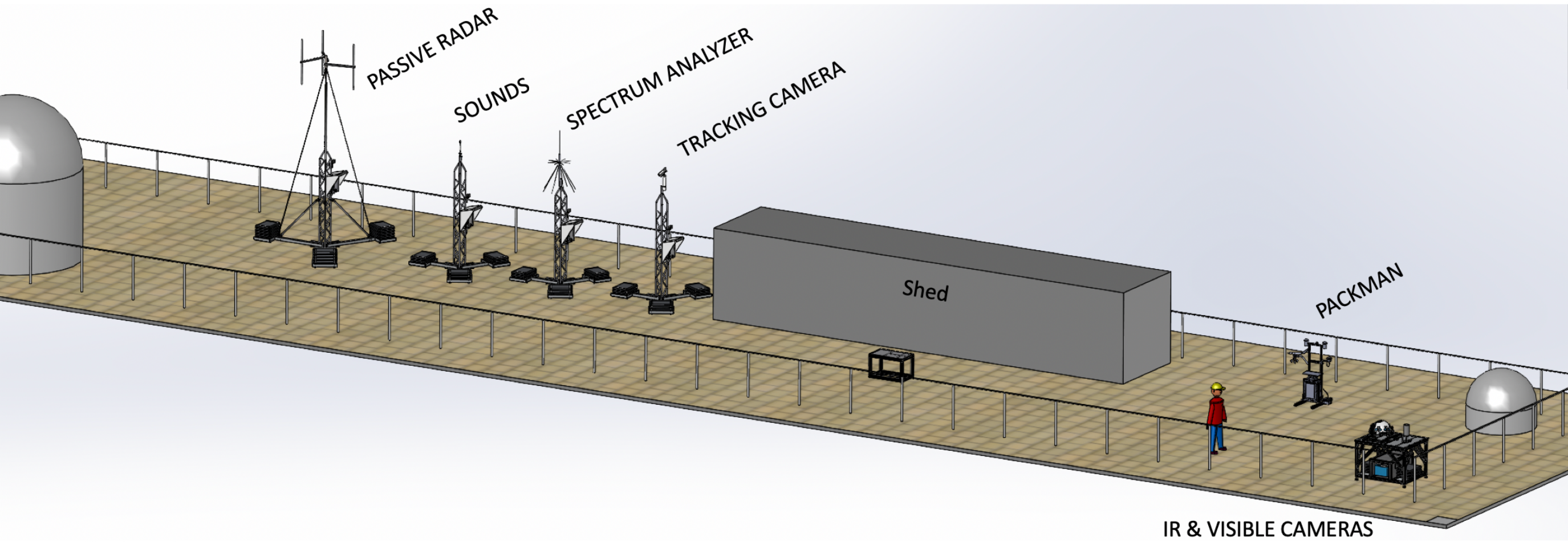}
    \caption{First all-sky Galileo Observatory for UAP on the roof of the Harvard College Observatory (Cambridge, MA).}
    \label{fig:fig1}
\end{figure}

\begin{figure}[htb!]
  \centering
  \includegraphics[width=\linewidth]{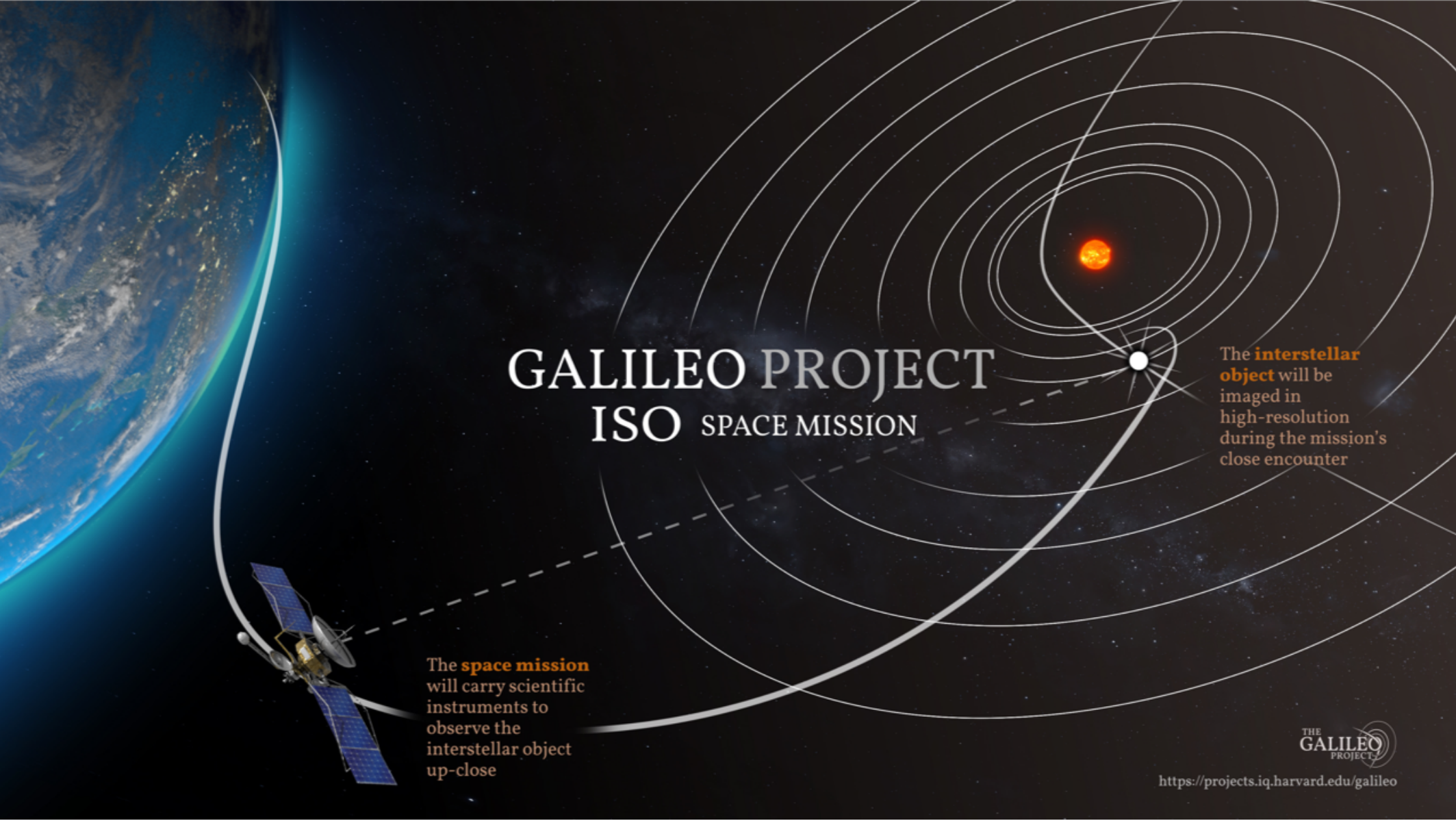}
    \caption{Space mission to intercept or rendezvous with the next `Oumuamua.}
    \label{fig:fig2}
\end{figure}

\begin{figure}[htb!]
  \centering
  \includegraphics[width=\linewidth]{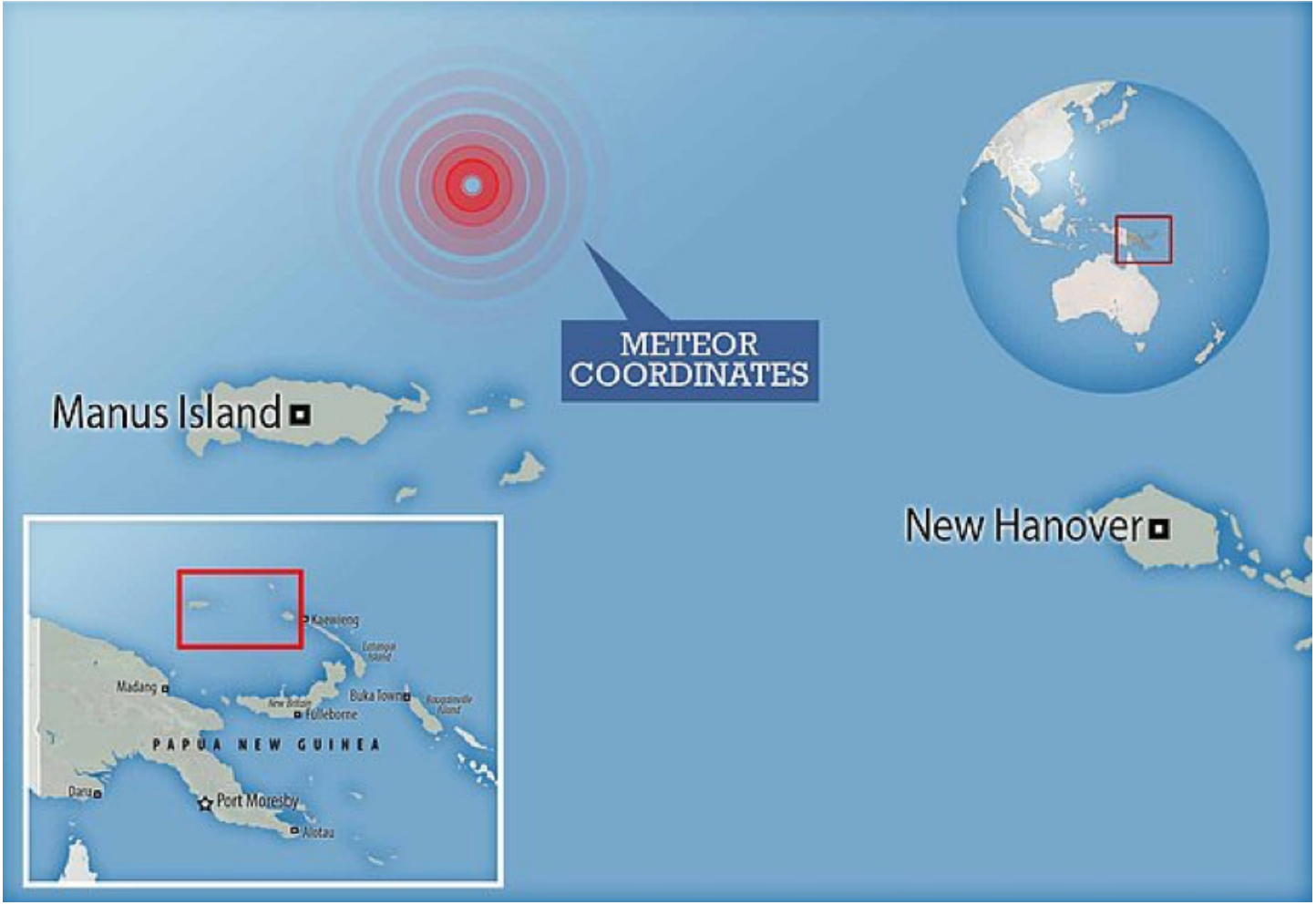}
    \caption{Location of CNEOS 2014-01-08 for the first Galileo Project ocean deep-water expedition.}
    \label{fig:fig3}
\end{figure}

The Galileo Project has three branches of activity~\citep{gpactivities}: 

\begin{enumerate}
    \item Constructing new telescope systems to infer the nature of Unidentified Aerial Phenomena (UAP), similar to those mentioned in the ODNI report~\citep{odnireport} to the US Congress (see Figure~\ref{fig:fig1}).
    \item Mining high-quality telescope data, e.g. from the Vera C. Rubin Observatory or from the Webb telescope to discover anomalous interstellar objects, and designing intercept or rendezvous space missions that that will identify the nature of interstellar objects that do not resemble comets or asteroids, like `Oumuamua~\citep{2021arXiv211015213L} (see Figure~\ref{fig:fig2}).
    \item Coordinating land or ocean expeditions to study the nature of interstellar meteors, like CNEOS 2014-01-08~\citep{2019arXiv190407224S} (see Figure~\ref{fig:fig3}).
\end{enumerate}

The non-profit Galileo Project has drawn a remarkable base of expert volunteers, from astrophysicists and other scientific researchers, to hardware and software engineers, to non-science investigators and generalists who volunteer their time and effort to the project in various ways.  The project brings together a broad community of members united by the agnostic pursuit of reliable and rigorous new scientific evidence through new telescopes, like the parallel, multi-modal and multi-spectral Galileo observatories - without prejudice. without prejudice. The project values the input of many different voices, and the rapid progress it has already made is a testament to its open approach. As different as the perspectives of the researchers and affiliates may be, every contributor to the Galileo Project is bound by three ground rules:

\begin{enumerate}
    \item The Galileo Project  is only interested in openly available scientific data and a transparent analysis of it.  Thus, classified (government-owned) information, which cannot be shared with all scientists, cannot be used. Such information would compromise the scope of the scientific research program of the Project, which is designed to acquire verifiable scientific data and provide transparent (open to peer review) analysis of this data. Like most physics experiments, the Galileo Project will work only with new data, collected from its own Galileo Observatories, which are under the full and exclusive control of Galileo research team members.
    \item The analysis of the data will be based solely on known physics and will not entertain fringe ideas about extensions to the standard model of physics. The data will be freely published and available for peer review as well as to the public, when such information is ready to be made available, but the scope of the research efforts will always remain in the realm of scientific hypotheses, tested through rigorous data collection and sound analysis. 
    \item To protect the quality of its scientific research, the Galileo research team will not publicize the details of its internal discussions or share the specifications of its experimental hardware or software before the work is finalized. The data or its analysis will be released through traditional, scientifically-accepted channels of publication, validated through the traditional peer-review process. The Project has no commercial interests.
\end{enumerate}

All members of the Galileo Project team, including researchers, advisors and affiliates, share these values and uphold the principles of open and rigorous science upon which the Galileo Project is founded.

The Galileo team developed a design of parallel, simultaneous multi-modal and multispectral imaging of UAPs, as well as an expedition to scoop the ocean floor near Papua New Guinea for the fragments of the first interstellar meteor, CNEOS 2014-01-08, and is designing a space mission to intercept or rendezvous unusual interstellar objects like `Oumuamua, to be identified in the future from the data pipeline of LSST on the Vera C. Rubin Observatory or other telescopes.

\section{First Year Studies}

The accompanying collection of papers celebrates the progress made in the first year of the Galileo Project (August 2021 through July 2022). The papers include an overview of the various branches of scientific investigation within the Galileo Project:

\begin{enumerate}
    \item Interstellar Objects.
    \item Interstellar Meteors.
    \item Unidentified Aerial Phenomena (UAP).
    \item UAP Computing and Processing Pipeline.
    \item UAP Wide Field Observations.
    \item UAP Acoustics.
    \item UAP Radar Network.
    \item UAP Satellite Data.
\end{enumerate}

The newly assembled Galileo Observatories are expected to collect and analyze data in the coming months and years. The Project team will reports its findings in peer reviewed journals and make the new data openly available to the scientific community after an initial verification period. 

\section{Expectations}

Extraterrestrial space archaeology~\citep{loeb_et_tech} is engaged with the search for relics of other technological civilizations~\citep{lingam_loeb_2021}. As argued by John von Neumann, the number of such objects could be extremely large if they are self-replicating~\citep{1980JBIS33251F}, a concept enabled by 3D printing and AI technologies. Physical artifacts might also carry messages, as envisioned by Ronald Bracewell~\citep{1960Natur186670B,1985AcAau121027F}. 

Searching for objects in space resembles a survey for plastic bottles in the ocean as they keep accumulating over time. The senders may not be alive when we find the relics. These circumstances are different from those encountered by the famous Drake equation~\citep{lingam_loeb_2021,DE}, which quantifies the likelihood of detecting radio signals from extraterrestrials. That case resembles a phone conversation in which the counterpart must be active when we listen. Not so in extraterrestrial archaeology.

What would be the substitute to Drake's equation for extraterrestrial archaeology in space? If our instruments survey a volume $V$, the number of objects we find in each snapshot would be~\citep{loeb_drake},
\begin{equation}
  N=n\times V  ~,
\end{equation}
where $n$ is the number of relics per unit volume. Suppose on the other hand that we have a fishing net of area $A$, like the atmosphere of the Earth when fishing meteors. In that case, the rate of new objects crossing the survey area per unit time is:
\begin{equation}
  R=n\times v\times A ~,
\end{equation}
where $v$ is the characteristic one-dimensional velocity of the relics along the direction perpendicular to that area. 

For life-seeking probes with maneuvering capabilities, the number density $n$ may be higher in the vicinity of habitable planets. Correspondingly, in the outskirts of planetary systems such probes are more likely to possess plunging orbits aimed radially towards the host star. In that case, the abundance of interstellar objects could be overestimated considerably by assuming an isotropic velocity distribution for detections near Earth.

Both $n$ and $v$ are likely to be functions of the size of the objects. NASA launched many more small spacecraft than large ones. In addition, the launch of faster objects increases the specific  energy requirements and may therefore be restricted to smaller objects that are more challenging to discover. Astronomical searches often target speeds of several tens of ${\rm km~s^{-1}}$ in the vicinity of Earth, as they are characteristic for asteroids or comets bound to the Sun. Advanced propulsion methods, such as light sails, could potentially reach the speed of light~\citep{GL15}, which is four orders of magnitude larger. Fast-moving objects might have been missed in past astronomical surveys, and should be considered in LSST data. Humanity's accomplishments thus far are modest. Over the past century, NASA launched five spacecraft that will reach interstellar space within tens of thousands of years: Voyager 1, Voyager 2, Pioneer 10, Pioneer 11 and New Horizons.

The detection threshold of surveys which rely on reflected sunlight sets the minimum size of a detectable object as a function its distances from the observer and the Sun. Moreover, comets are more easily detectable than non-evaporating objects, because their tail of gas and dust reflects sunlight beyond the extent of their nucleus. Meteors, on the other hand, are found through the fireball they produce as they disintegrate by friction with air in the Earth's atmosphere. This makes meteors detectable at object sizes that are orders of magnitude smaller than space objects. For example, CNEOS 2014-01-08 was merely $\sim 0.5$m~\citep{2022RNAAS681S} in size whereas a sunlight-reflecting object like `Oumuamua was detectable within the orbit of the Earth around the Sun because its size was $\sim 100-200$m~\citep{Trilling}. The nucleus of the comet Borisov was $\sim 200$-$500$m in size~\citep{Jewitt}, and its evaporation made the comet detectable even farther because of its larger tail. NASA never launched spacecraft as big as `Oumuamua. Interstellar objects like CNEOS 2014-0108 are a million times more abundant than `Oumuamua near Earth, but they were not detectable by the Pan STARRS survey which discovered `Oumuamua.

Electromagnetic (e.g. radio or laser) signals escape from the Milky-Way galaxy and reach cosmological scales over billions of years. However, chemical rockets are generically propelled to speeds of tens of ${\rm km~s^{-1}}$, which are an order of magnitude smaller than the escape speed from the Milky Way. Coincidentally, this speed is sufficient for escape from the habitable zone of a Sun-like star when combined with the orbital speed of a parent Earth-like planet. Moreover, this speed is comparable to the velocity dispersion of stars in the disk of the Milky Way. As a result, interstellar chemical rockets remain gravitationally confined to the Milky-Way disk within roughly the same vertical scale-height as their parent stars (hundreds of parsecs).  The cumulative abundance of such objects would be set by an integral over their production history per star following the star formation history of the Milky Way. 

Just like terrestrial monuments, space artifacts provide evidence for past civilizations. They continue to exist in the Milky Way even if the technological era of their senders lasted for a short window of time relative to the age of the Galaxy, such that none of these senders transmits radio signals at present.

In difference from electromagnetic signals, the abundance of interstellar artifacts which are gravitationally-bound to the Milky-Way disk would grow over cosmic time. The abundance of small objects is likely to be much larger than large objects, in part because some of them may represent fragments generated by the destruction of larger objects.

Based on the cosmic star formation history~\citep{Madau}, most stars formed billions of years before the Sun, allowing sufficient time for chemical rockets to disperse through the Milky-Way disk if civilizations like ours emerged with the same time lag after the formation of other Sun-like stars. But even if one civilization had launched self-replicating probes, the abundance of artificial probes can be very high within the entire Milky-Way galaxy.

This all assumes that we are searching. But there is a probability, $O$, that some scientists might behave like an ostrich and avoid the search for interstellar objects of technological origin altogether. For example, LSST data could be analyzed only by fitting orbits bound to the Sun. Similarly, funding agencies could decide not to engage in any search that deviates from the beaten path. The final equations are therefore:
\begin{equation}
  N=n\times V\times (1-O) ~,
\end{equation}
and,
\begin{equation}
  R=n\times v\times A\times (1-O) ~.
\end{equation}
The likelihood of us finding extraterrestrial technological objects depends on us being open-minded and willing to search for them and not just on whether the extraterrestrials had sent them.

An interstellar object of future interest could be studied in great detail by the James Webb Space Telescope (JWST)~\citep{ST} as it passes nearby. Since JWST is located a million miles away from Earth at the second Lagrange Point L2, it would observe the object from a completely different direction than telescopes on Earth. This would allow us to map the three-dimensional trajectory of the object to exquisite precision and determine any forces~\citep{Micheli} acting on it in addition to the Sun's gravity. Moreover, JWST would be able to detect the spectrum of infrared emission and reflected sunlight from the object, allowing JWST to potentially infer the composition of its surface. 

To gain even better evidence, it would be beneficial to bring a camera close to the object on its approach, as planned by the Galileo Project. Better yet would be to land on the object and take a sample from it back to Earth as the OSIRIS-REx mission did with the asteroid Bennu~\citep{osirisrex}. 

A different opportunity to put our hands on material from such an object would be to examine the remains of interstellar meteors that are of technological origin~\citep{loeb_first_iso}. Whereas a space mission often requires billions of dollars in funding, the latter approach is a thousand time lower in cost.

\section{Concluding Remarks}

If the Galileo Project’s search will find indisputable evidence for an object that is not natural or human-made, then this finding would be a teaching moment for humanity. It might provide a simple answer to Fermi’s paradox~\citep{lingam_loeb_2021}: ``where is everybody?'', in the form of: ``right here.'' Scientists have been searching for sixty years for radio signals from planets around distant stars~\citep{lingam_loeb_2021,etsearch}, but they neglected to check systematically for interstellar objects in our backyard.

The second branch of the Galileo Project involves the design of a space mission to intercept or rendezvous with unusual interstellar objects like `Oumuamua, in the spirit of NASA’s OSIRIS-REx mission — which landed on the asteroid Bennu, or ESA’s plan for a future Comet Interceptor~\citep{esa} — which is limited in its maneuvering speed. The Galileo Project will develop software that will identify interstellar objects that do not resemble familiar asteroids or comets from the Solar system. This software will be applied to the LSST data pipeline.

Finally, a third branch of the Project involves a plan for an expedition to retrieve fragments from the first interstellar meteor CNEOS 2014–01–08~\citep{loeb_iso_iron} from the ocean floor near Papua New Guinea.

The outcome of scientific research cannot be forecasted. The Astronomy Decadal Survey in 2010~\citep{NAP12951} did not anticipate the main discoveries of last decade, such as the first detection of gravitational waves in 2015~\citep{gravitational_waves}, the discovery of the interstellar object - `Oumuamua in 2017, and the imaging of the black hole in M87 in 2019~\citep{messier}. These items were not even listed as high-level priorities in astrophysics a decade ago. Here’s hoping that the findings of the Galileo Project will be the highlight of the next decade in astronomy.

The responsible approach of scientists should be to attend to new evidence as unusual as it might be, and adapt to its implications irrespective of how challenging they are. 

What we regard as ``ordinary'' are things we are used to seeing. Such things include birds in the sky. But digging deeper into the nature of ordinary matters suggests that they are rather extraordinary. Humans were only able to imitate birds with the Wright brothers’ first flight in 1903. Similarly, what we regard as ``extraordinary claims'' is often based on societal conventions. We had been investing billions of dollars in the search for the nature of dark matter whose existence was initially doubted for four decades after Fritz Zwicky first proposed its existence in 1933~\citep{Zwicky}; yet we still allocate minimal funds to the scientific study of UAP. As a result, the lack of ``extraordinary evidence'' is often self-inflicted ignorance. We have little chance of finding extraordinary evidence for our cosmic neighbors unless we look through our windows and actively engage in the search for anomalous objects, including `letters' in our mailbox of the solar system. By engaging in the search, we might figure out the nature of UAP before we understand dark matter, if we would only be brave enough to collect and analyze UAP data publicly, based on the scientific method.

The instruments developed by the Galileo Project represents a brand-new observatory design with unprecedented capabilities. As its “Lego pieces” are put together, our hearts fill with appreciation for the professional quality of the Galileo team members. In the years to come, we will harvest new knowledge from these new telescope systems.

These Galileo observatories are the new eyes and the computer system attached to them is the new brain of the Galileo Project. Watching the sky through new observatories is our best way to find out whether we have neighbors.  What we do with the answer depends on the details it entails. As Robert Frost noted in his poem ``The Road Not Taken'':``Two roads diverged in a yellow wood… I took the one less traveled by, And that has made all the difference.''~\citep{frost1951road}. There is a great advantage to taking the road not taken. If there is any low hanging fruit along that path, the Galileo Project will harvest it.

\bigskip
\bigskip
\bigskip
\noindent
{\bf ACKNOWLEDGEMENTS.} The Galileo Project is supported by generous donations from Eugene Jhong, Vinny Jain, Teddy Jones, Eric Keto, Laukien Science Foundation, Joerg Laukien, William A. Linton, Adnan Sen and The Brinson Foundation. Our special gratitude to Wes Watters and Richard Cloete, the {\it Laukien-`Oumuamua Postdoctoral Fellow} of the Galileo Project, for insightful comments on the manuscript.
\bigskip
\bigskip
\bigskip

\bibliography{references}{}
\bibliographystyle{aasjournal}



\end{document}